# On Timing Model Extraction and Hierarchical Statistical Timing Analysis

Bing Li, Ning Chen, Yang Xu, Ulf Schlichtmann, *Member, IEEE*

*Abstract*—In this paper, we investigate the challenges to apply Statistical Static Timing Analysis (SSTA) in hierarchical design flow, where modules supplied by IP vendors are used to hide design details for IP protection and to reduce the complexity of design and verification. For the three basic circuit types—combinational, flip-flop-based and latch-controlled—we propose methods to extract timing models which contain interfacing as well as compressed internal constraints. Using these compact timing models the runtime of full-chip timing analysis can be reduced, while circuit details from IP vendors are not exposed. We also propose a method to reconstruct the correlation between modules during full-chip timing analysis. This correlation can not be incorporated into timing models because it depends on the layout of the corresponding modules in the chip. In addition, we investigate how to apply the extracted timing models with the reconstructed correlation to evaluate the performance of the complete design. Experiments demonstrate that using the extracted timing models and reconstructed correlation full-chip timing analysis can be several times faster than applying the flattened circuit directly, while the accuracy of statistical timing analysis is still well maintained.

## I. Introduction

IN Statistical timing analysis process variations are modeled by random variables to overcome the pessimism from traditional worst-case timing analysis. According to the distributions of random variables and the mapping functions from process parameters to delays, statistical timing analysis methods introduced in recent years can roughly be classified as first-order methods [1]–[3], second-order methods [4]–[8] and methods handling non-Gaussian variations and nonlinear mapping functions [9], [10]. For example, the method in [8] introduces the orthogonal polynomials to characterize the stochastic process from transistors to logic gates and to IP blocks with very accurate results and high efficiency. Moreover it provides a mechanism for statistical characterization of cell libraries. In these methods, all timing properties, such as path delays, slacks and the minimum clock period, become random variables, in contrast to the fixed values in traditional worst-case timing analysis. This change forces us to reinvestigate many steps of the IC design flow when adopting statistical timing analysis, because the design of digital circuits heavily depends on the result of timing analysis.

In this paper we focus on the application of statistical timing analysis to the hierarchical design flow, where predesigned functional blocks, called modules in the following, are integrated into complex chips to reduce the cost of design and verification. In this design flow, the internal timing constraints of a module need not to be exposed to users completely and therefore can be compressed into a much simpler form. The timing constraints from interfacing logic should also be kept in the precharacterized timing models. These interfacing constraints may be combinational timing paths from inputs to outputs of the module; or they may be setup time constraints at latches that can be reached from inputs due to latch transparency. These extracted timing constraints together form the timing model of the module. With the compact timing models, the efficiency and capacity of full-chip analysis can be improved significantly. Because very little circuit details are contained in timing models, this hierarchical design flow has the additional advantage to protect intellectual property.

Several methods have been proposed to address the arising challenges in the hierarchical design flow when process variations are considered. The pioneering work in [11] extends the extracted timing model (ETM) in [12] to incorporate statistical delays by mapping the path delays in the constraints into functions of global and local variations, and sample-based SPICE simulations are used to identify the sensitivities of the random variables modeling components of process variations. It also provides a solution for reconstructing the correlation between modules by solving linear systems created from transformation matrices. Besides [11], the method in [13] applies graph compression and sensitivity reduction to reduce model size; and the method in [14] groups transistors in custom designs into large blocks, whose pin-to-pin delays are characterized using SPICE simulations. These methods contribute to the simplification of statistical timing models, but many circuit-level challenges, for example, the transparency of latches, still require further research work.

In searching for solutions to incorporate process variations into the hierarchical design flow, naturally we may explore existing timing model extraction methods proposed for static timing analysis, where delays as well as other timing properties are modeled with fixed values. For combinational circuits, the methods [12], [15] extract the critical paths between inputs and outputs directly, resulting in black-box timing models [15]. In [16], special structures and delay patterns are processed to

This work was supported in part by the German Research Foundation as part of the Transregional Collaborative Research Center, Invasive Computing (SFB/TR 89).

Bing Li, Ning Chen and Ulf Schlichtmann are with the Institute for Electronic Design Automation, Technische Universität München, Munich 80333, Germany (e-mail: b.li@tum.de; ning.chen@tum.de; ulf.schlichtmann@tum.de).

Yang Xu is with Intel Mobile Communications, Neubiberg 85579, Germany (e-mail: yang.a.xu@intel.com).

compress timing models. In [17], the biclique-star replacement technique is proposed to minimize the number of edges. From these methods, only the basic graph transformations in [16] can be applied to statistical timing model extraction directly. The other techniques are infeasible since they rely on certain delay patterns. For circuits using edge-triggered flip-flops, the methods in [12], [15] extract the timing arcs from inputs to the first level of flip-flops and from the internal flip-flops to the outputs. These methods can be applied in statistical timing analysis directly, though the delay and constraint arcs have weights determined not by a single critical path but by the maximum of many path delays statistically. For circuits using level-sensitive latches, timing model extraction becomes more complex, because delay and constraint paths may span multiple stages of latches due to latch transparency. In [15] all latches are retained in timing models to allow arbitrary time borrowing. In [18], [19] delays and constraints across latches are extracted assuming given clock waveforms. The delay comparison used in these methods, however, prevents their application in handling process variations. In [12], [20] a maximum level of latch transparency is assumed during model extraction. If the level of latch transparency of a module implemented in a design exceeds that assumed during model extraction, a new model should be recharacterized by the IP vendor. This process makes design reuse more difficult and increases the interactions between IP vendors and chip designers. However, the basic idea of constraining the level of transparency can be borrowed into statistical timing model extraction, while new formulations and algorithms are necessitated to reduce the risk of recharacterization and to improve the modeling accuracy.

For all the circuit types described above, false paths can be excluded during timing model extraction with the methods [21], [22] by considering input vectors, so that the pessimism from topological delay computation can be alleviated; and crosstalk information can also be included in the timing models [23]. These methods may be combined with the methods for individual circuit types to improve modeling accuracy in static as well as statistical timing analysis.

In this paper, we investigate the complete flow of hierarchical statistical timing analysis. Our main contributions are as follows:

- For combinational circuits, we propose a method to extract statistical timing models using noncritical edge removal and structural transformations. Instead of using the maximum delays between inputs and outputs directly, the proposed method compresses timing graphs constructed from the original circuits, leading to very compact combinational timing models.
- For sequential circuits using edge-triggered flip-flops, we adapt the ETM model in [12] to capture the statistical interfacing constraints. The constraints between internal flip-flops are compressed and explicitly included for full-chip analysis. These internal constraints are implicitly specified in static timing models but not handled by existing methods for statistical timing model extraction.
- For sequential circuits using level-sensitive latches, we introduce two constraint sets with which a lower bound and an upper bound of the statistical minimum clock period can be computed. We analyze the relation between the given transparency level and the distance of the two bounds quantitatively so that tradeoff between model size and accuracy is possible.
- For full-chip timing analysis, we improve the correlation handling method in [11] to guarantee a unique variable substitution. We also address the challenges in the application of the extracted timing models, for example, the loops formed between modules due to latch transparency.

Compared with the work in [11] we focus on the identification of critical gates in combinational timing models. The delays of these critical gates can be mapped into functions of global and local variations using the method in [11] effectively. We also propose techniques to extract timing constraints between and across sequential cells. In addition, we extend the correlation reconstruction method in [11] to provide a unique solution. Combining the method in this paper and that in [11], we can form a systematic framework for hierarchical statistical timing analysis.

The rest of this paper is organized as follows. In Section II we give a short introduction on delay representation in statistical timing analysis. The correlation reconstruction in Section IV-A is based on this general delay representation. In Section II, we also give an overview of the different types of constraints which may appear in timing models. In Section III we explain the proposed methods to generate statistical timing models for the three basic circuit types—combinational, flip-flop-based and latch-controlled. In Section IV we explain the correlation reconstruction and the application of the statistical timing models in full-chip timing analysis. We show experimental results in Section V and conclude this paper in Section VI.

## II. Preliminaries

In this section we explain the representation of statistical gate delays. The correlation between these delays are maintained by sharing the same set of independent random variables from decomposition. Thereafter, we give an overview of the different types of constraints which may be included in the extracted timing models.

### A. Statistical Gate Delays

Under process variations, a process parameter is represented using a random variable. According to the spatial characteristics, this variable can be partitioned into different parts: global variation, local variation and independent variation. Consequently, a parameter $p$ can be written as

$$p = p_0 + p_g + p_l + p_r \tag{1}$$

where $p_0$ is the nominal value of the parameter; $p_g$ models the global variation which is shared by all devices on the die; $p_l$ models the local variation, which is specific to each device and correlated with each other; $p_r$ is an independent variable modeling the purely random effect in the manufacturing processes.

Local variations are specific to devices and their correlation, also called spatial correlation, depends on the distance between devices on the die [24]. Owing to the large number of devices in the circuit, it is impractical to assign a random variable to each device to represent the local variation. Because the correlation does not change drastically along distance [25], the die area is usually partitioned into a grid and all devices in a grid cell share the same variable, as proposed in [1]. An example of this partition is illustrated in Fig. 1, where each small square is called a grid cell. Note that the die partition is not necessarily uniform as in [1], since large grid cells can be applied for the area where correlation changes slowly [26]. Furthermore, the shapes of the grid cells may not always be rectangular. For example, hexagons are used in [27] to partition the die area.

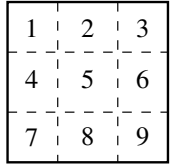

Fig. 1. Die partition modeling local variations.

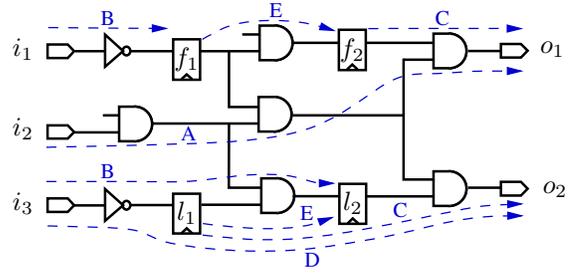

Fig. 2. Types of timing constraints.

For the $i$th grid cell, a random variable $p_{l,i}, i \in \{1, 2, \ldots, n\}$ is assigned to model the local variation. The random variables for all the $n$ grid cells are denoted by the vector $\mathbf{p}_l$. To reduce the computational complexity during statistical timing analysis, the correlated variables $\mathbf{p}_l$ are decomposed so that each $p_{l,i}$ is represented by a linear combination of the same set of independent random variables, using principle component analysis (PCA) [1] or independent component analysis (ICA) [10], as

$$\mathbf{p}_l = \mathbf{A}\mathbf{x} \quad (2)$$

where $\mathbf{x} = [x_1, x_2, \ldots, x_n]^T$ is a set of independent random variables and the transformation matrix $\mathbf{A}$ is orthogonal, so that $\mathbf{A}^{-1} = \mathbf{A}^T$. The random variable $p_{l,i}$ assigned to the $i$th grid cell is an element of $\mathbf{p}_l$, so that it can be expressed as a linear combination of $\mathbf{x}$, as

$$p_{l,i} = \mathbf{k}\mathbf{x} \quad (3)$$

where $\mathbf{k}$ is the row vector formed by the $i$th row of $\mathbf{A}$.

To explain the modeling of gate delay $d$, we use only one process parameter $p$ as example. The relation between $d$ and $p$ can be expressed as a function $f$. Therefore, the delay $d$ of a gate in the $i$th grid cell can be written as

$$d = f(p) = f(p_0 + p_g + p_{l,i} + p_r) = f(p_0 + p_g + \mathbf{k}\mathbf{x} + p_r) \quad (4)$$

For simplicity, we write $d$ as

$$d = g(\mathbf{k}\mathbf{x}) \quad (5)$$

where the nominal value $p_0$ and the variables $p_g$ and $p_r$ in (4) are implicitly included in the function $g$, for convenience of the discussion in Section IV-A.

In statistical timing analysis, the arrival times, slacks and path delays are usually represented in a similar form of (5), so that the arrival time propagation can be performed recursively with the same maximum and sum operations. In Section IV-A we will explain a method to reconstruct the correlation between modules based on the general form (5).

### B. Constraints in Timing Models

A timing model for hierarchical analysis contains the interfacing timing constraints and the abstracted internal timing constraints of a module. The interfacing constraints are from the circuit components that interact with other modules. The internal timing constraints are between the components that are not affected by the application context. Therefore these constraints can be abstracted into a much simpler form to reduce the size of the timing model. In this section, we explain five basic types of timing constraints, in which type A–D are interfacing constraints and type E is the internal constraint. A circuit segment in Fig. 2 illustrates examples of these constraints, where $f_1$ and $f_2$ are edge-triggered flip-flops; $l_1$ and $l_2$ are level-sensitive latches. For simplicity, we only discuss the constraints from setup times in this paper and the constraints from hold times can be deduced similarly.

The constraints of type A are from combinational circuits. In such circuits, a signal from an input can traverse through combinational paths and reach outputs. For example, the signal from input $i_2$ can reach $o_1$ and $o_2$ without being interrupted by flip-flops or latches. To guarantee that the arrival time at any output can be computed correctly when different arrival times of the signals are applied to the inputs, the extracted timing model should contain the same maximum delays between any pairs of input and output. To retain these timing constraints, the black-box methods, for example, the ETM model in [12], save these maximum delays in the timing model directly. For a combinational module with $m$ inputs and $n$ outputs, the number of these constraints may reach $m \times n$. When $m$ and $n$ are large, the black-box model of the module consumes much memory and its efficiency degenerates. In Section III-A, we will explain a method based on noncritical edge removal and graph transformations to overcome the disadvantage of the black-box methods.

The constraints of type B are extracted for the flip-flops and latches that are reachable from inputs. For example, the signal from $i_1$ can reach $f_1$ across a combinational path and the arrival time should meet the setup time constraint of $f_1$. For modules using flip-flops, these constraints can simply be computed by propagating arrival times from individual inputs, for example, using [28]. For modules with level-sensitive latches it is more difficult to extract these constraints, because signals can pass through many latch stages transparently. For example, the timing constraint from $i_3$ to $l_2$ should also be extracted if $l_1$ has a chance to be transparent. The major challenges in extracting these constraints are that the level of latch transparency is unknown during timing model extraction and that statistical delays make these transparency conditions probabilistic.

The constraints of type C are the delays from internal flip-flops or latches to outputs. For example, the signal propagated from the flip-flop $f_2$ reaches $o_1$ with the maximum delay from $f_2$ to $o_1$. This signal may continue the propagation in the following modules and eventually be checked against the setup time constraint at a flip-flop or latch. Therefore the maximum delays from internal flip-flops or latches to outputs should be included in the timing model. The difficulty of extracting these timing constraints still exists for latch-controlled circuits, because the signals from internal latches may reach outputs with certain probabilities. For example, the signal from $l_1$ may pass through $l_2$ and reach $o_2$ if $l_2$ is transparent.

The constraints of type D are the maximum delays from inputs to outputs across latches, which may be transparent according to the arrival times of the signals at the inputs. For example, the signal from $i_3$ may reach $o_2$ if $l_1$ and $l_2$ both are transparent. However, to evaluate these transparencies directly, or pessimistically, may be much time-consuming and result in large timing models.

The last type of constraints, type E, is the compressed internal timing constraints between flip-flops or latches. This is actually the result of timing analysis of the module without considering the circuit parts connected with inputs and outputs. For circuits using flip-flops, these constraints can be computed by a standard statistical timing engine very efficiently and represented by the minimum clock period for the module. This constraint must explicitly be included in statistical timing models, because the minimum clock period is in the form of a random variable, and it should be verified together with the interfacing constraints between modules. For circuits with transparent latches, the internal constraints can be extracted using existing statistical timing engines, for example, [29]–[31], if the clock period is given. Otherwise the method [32], which we use in this paper, can be used to extract more flexible timing constraints.

## III. STATISTICAL TIMING MODEL EXTRACTION

In this section we propose several methods to extract the statistical timing constraints described in Section II-B from the three basic circuit types—combinational, flip-flop-based and latch-controlled—individually. In practice these circuit types may coexist in a large design and the proposed methods can be combined together to generate timing models. In the following, all variables, except those which we clearly stated as constants, are random and represented in the statistical form (5). The maximum and sum operations mentioned in this paper are very general and do not rely on any properties of the underlying distributions of the random variables, so that existing statistical timing engines, for example, [1]–[10], can be applied directly to handle random variables of different forms.

### A. Statistical Timing Models for Combinational Circuits

Statistical timing constraints from combinational circuits are all of type A described in Section II-B and illustrated in Fig. 2. The extracted timing model for a combinational module should have the same maximum delays between any pairs of input and output. In the proposed method, we first remove the delay edges that do not have a significant effect on any input-output delays. Thereafter, we apply two structural transformations to reduce the size of the timing model further.

We explain the basic idea of noncritical edge removal using an example illustrated in Fig. 3(a). In this example, the delay structure of the module is described by a *timing*

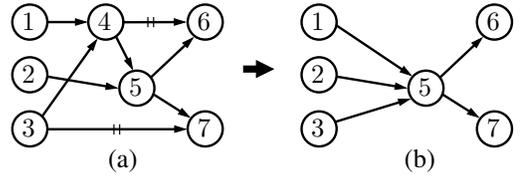

Fig. 3. Concept of noncritical delay removal. (a) Original timing graph. (b) Timing model.

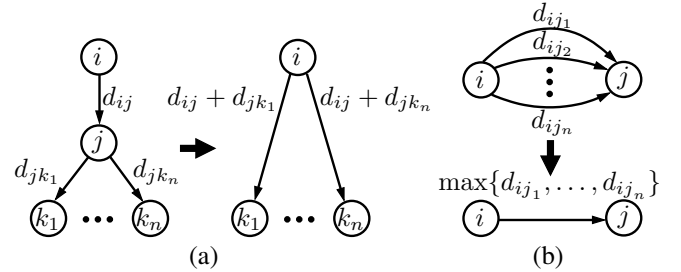

Fig. 4. Basic graph transformations. (a) Serial. (b) Parallel.

*graph*, where edges represent gate delays and nodes represent circuit inputs, outputs and connections. Between an input and an output, usually there are many combinational paths, and the maximum delay is determined by the critical path(s). In statistical timing analysis, all delays are modeled by random variables, so that each path has either a large or a small probability to be critical. To extract the timing model for a combinational module, we compute the probabilities that the edges appear on the critical paths between input-output pairs using the concept of *criticality* [33]–[36]. Edges with small probabilities to affect input-output delays are then deleted. For example, the two edges marked by ∥ in Fig. 3(a) can be removed if they have small criticalities. With a low threshold of criticalities in removing edges, this technique reduces the size of the timing model with hopefully minimal effect on accuracy.

The criticality of a delay edge is the probability that it appears on critical paths after manufacturing [33]. For a given circuit input $p$ and output $q$, the criticality $c_{ij}^{pq}$ of an edge between node $i$ and $j$ is the probability that the edge is on a critical path between input $p$ and output $q$. This criticality can be computed using existing methods efficiently, for example, [33]–[36]. We then compute the maximum criticality $c_{ij}^m$ from all $c_{ij}^{pq}$ as

$$c_{ij}^m = \max_{p,q}\{c_{ij}^{pq}\} \quad (6)$$

where the maximum operation is performed over all input-output pairs. If the resulting $c_{ij}^m$ of a delay edge is less than a predefined small threshold $\delta_c$ approximating 0, we remove this edge to compress the timing model, because it has only a small probability to affect input-output delays and therefore the model accuracy.

After removing edges with small criticalities, we apply the basic transformations used in [15], [16], [37], [38] to compress the timing model further. The two transformations are illustrated in Fig. 4. The *serial transformation* in Fig. 4(a) reduces the number of edges by 1. The *parallel transformation* in Fig. 4(b) merges all the parallel edges between $i$ and $j$ into one edge. These two transformations retain the maximum delay from $i$ to $k_1, k_2, \ldots, k_n$ in the serial case and to $j$ in the parallel case, so that they do not affect input-output delays of the timing model. As an example, after removing the two noncritical edges in Fig. 3(a), we can apply the serial transformation to the pattern including nodes 1, 3, 4, and 5 to generate the timing model shown in Fig. 3(b). Although no examples of parallel edges are shown in Fig. 3, these edges may also exist after applying the technique of noncritical edge removal, because they can not dominate each other definitely due to process variations. The final timing graph is used as the timing model of the module, which typically contains much fewer nodes and edges than the original timing graph.

### B. Statistical Timing Models for Flip-flop-based Circuits

Statistical timing model extraction for a flip-flop-based module is very similar to the ETM method for static timing models in [12]. The first type of constraints is type B in Fig. 2, extracting the timing constraints at the flip-flops that can be reached by a signal from an input across combinational paths. From an input $i$, usually there is more than one flip-flop that can be reached from $i$. Therefore, the timing constraint for input $i$ is

$$\max_j \{\tilde{a}_i + d_{ij} + s_j\} \leq T \iff \quad (7)$$
$$\tilde{a}_i + \max_j \{d_{ij} + s_j\} \leq T \iff \quad (8)$$
$$\tilde{a}_i + c_i \leq T \quad (9)$$

where $\tilde{a}_i$ is the arrival time at the input $i$ and unknown during timing model extraction; $T$ is the clock period; and $s_j$ is the setup time of flip-flop $j$; and $d_{ij}$ is the maximum delay from input $i$ to flip-flop $j$. The max operation is performed over all the flip-flops reachable from $i$. In statistical timing analysis, $c_i$ can be computed efficiently using [28] with statistical maximum and sum operations and is kept in the timing model to represent the timing constraint (9) for the input $i$.

Similar to the timing constraint for input $i$, the delays from internal flip-flops to each output should also be computed. These constraints are of type C in Fig. 2 and are needed to compute the arrival times at the inputs of the following modules. We keep a random variable $d_j$ for each output $j$ in the timing model to represent the maximum delay to $j$ and $d_j$ can easily be computed by propagating arrival times from flip-flops to outputs.

The timing constraints for inputs and outputs have the same definitions as those in the ETM model [12], except that $c_i$ and $d_j$ are computed from all combinational paths instead of only the critical paths in static timing analysis. Another difference between the statistical model and the ETM model is that the timing constraints between internal flip-flops should be explicitly verified during full-chip analysis. Assuming that the maximum combinational delay between flip-flop $i$ and $j$ is $d_{ij}$ and the setup time of $j$ is $s_j$, these constraints can be written into a simple form, as

$$\max_{i,j} \{d_{ij} + s_j\} \leq T \iff T_m \leq T. \quad (10)$$

$T_m$ is actually the minimum clock period of the module without considering inputs and outputs, and can be computed very efficiently using a standard statistical timing engine. For static timing models, this internal constraint is implicitly assumed as the fact that each module must meet the specifi-

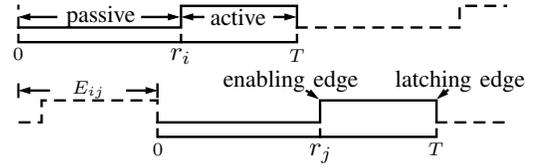

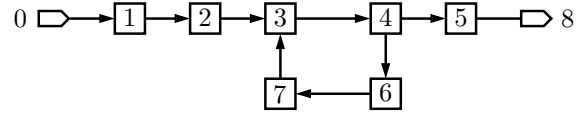

Fig. 5. Clock phases.

```
0 ▷──▶ 1 ──▶ 2 ──▶ 3 ──▶ 4 ──▶ 5 ──▷ 8
                   ▲      │
                   │      ▼
                   7 ◀── 6
```

Fig. 6. Latch graph.

cation of the clock period. In statistical timing analysis, the constraint (10) contains the yield information of each module itself. In addition, the variables $T_m$ from different modules are correlated. Therefore, these constraints must explicitly be included in computing the statistical minimum clock period during full-chip timing analysis. With $T_m$ as well as $c_i$ and $d_j$, the timing model for a flip-flop-based module only contains $m + n + 1$ random variables, where $m$ is the number of inputs and $n$ the number of outputs.

### C. Statistical Timing Models for Latch-controlled Circuits

Timing constraints in latch-controlled circuits are similar to those in flip-flop-based circuits but are more complex due to latch transparency. These constraints can be handled separately as: timing constraints between latches, timing constraints from primary inputs, and timing constraints from internal latches to primary outputs.

*1) Latch transparency and basic timing constraints*

The challenges of timing model extraction for latch-controlled circuits come from the fact that latches allow signals to pass transparently. Fig. 5 shows a two-phase clock scheme. If a signal reaches a latch with its clock in the active period, the signal can propagate across the latch transparently; otherwise, the signal must wait until the enabling edge to start the propagation. Owing to latch transparency timing constraints generated from the paths across multiple latch stages should also be considered. The challenges introduced by this characteristic in extracting timing constraints can be explained using a simple example shown in Fig. 6, where latches are represented with squares and combinational delays between latches with arrows. This structure is called *latch graph* in the following. In this example, if latch 1 is in transparency, the signal from the input may reach latch 2. Similarly, the signal from latch 4 in Fig. 6 can also reach the output if latch 5 is in transparency. The level of latch transparency from inputs is determined by the arrival times of the signals at the inputs of the module. But these arrival times are unavailable during timing model extraction. Even worse, in statistical timing analysis, all delays are random variables so that the level of transparency is also statistical. We solve this problem of transparency by allowing a fixed level of transparency. With this assumption we extract a lower-bound constraint set and an upper-bound constraint set. The numbers of constraints in these sets are then reduced using statistical dominance [39].

Consider the clock scheme in Fig. 5 for latch $i$ and $j$. If latch $i$ is in transparency, the arrival time at $j$ is computed by

$$a_j = a_i + d_{ij} - E_{ij} \quad (11)$$

where $a_i$ and $a_j$ are the arrival times defined with respect to the start time of the clock periods, respectively. $E_{ij}$ denotes the time difference between the two clock phases, and is called *phase shift* in the following. $d_{ij}$ is the maximum delay of combinational paths from $i$ to $j$. The positions of the rising clock edges are denoted by $r_i$ and $r_j$ in the clock phases. According to [32], [40], [41], a signal starting from the enabling clock edge of a latch after reset may reach any latch through a path of transparent latches, and the arrival time of the signal must meet the setup time constraint at the ending latch. Such a constraint of a path with $n_l$ latches is written as

$$a_j = r_i + \sum_{k=1}^{n_l-1}(d_{k,k+1} - E_{k,k+1}) \leq T - s_j \quad (12)$$

where the path starts from latch $i$ indexed by 1 and ends at latch $j$ indexed by $n_l$. The commas in $d_{k,k+1}$ and $E_{k,k+1}$ are used to avoid ambiguity in the notation. Examples of the constraint (12) are the cases from latch 1 to 2 in Fig. 6 as $r_1 + d_{1,2} - E_{1,2} \leq T - s_2$ and from latch 2 to 3 as $r_2 + d_{2,3} - E_{2,3} \leq T - s_3$.

*2) Timing constraints between latches*

The first timing constraint, type E between $l_1$ and $l_2$ in Fig. 2, of a latch-controlled circuit is similar to (10) for a flip-flop-based circuit. To guarantee the proper behavior of the module, all the timing constraints (12) between latches should be extracted. This is simply the task of statistical timing analysis considering transparent latches, though the computation is much more complex than that of the algorithm to extract the minimum clock period (10) for a flip-flop-based circuit, due to the large number of paths across latches. Existing timing engines [29]–[31] solve this problem using graph traversal or iterations when the clock period is given. For a more flexible constraint, the method in [32], which relies on reduced iterations and graph transformations, can be used to extract a statistical minimum clock period, so that the extracted timing model can be applied with different clock periods. This characteristic increases the flexibility of the timing model and reduces the interaction between IP users and vendors, so that we use this method for timing model extraction in this paper. Similar to (10), all the timing constraints (12) between latches are merged into a simple form in [32], as

$$T_m \leq T \quad (13)$$

where $T_m$ is the statistical minimum clock period of the module. The proposed method in this paper, however, does not depend on the method in [32], because we only assume that the statistical minimum clock period of a latch-controlled circuit can be computed. Any algorithms that solve this problem can directly be used to replace [32] in the proposed method.

*3) Concept of timing constraints from primary inputs*

The timing constraints of the second type that should be included in the timing model, type B in Fig. 2, come from the timing paths originate from circuit inputs. Because of latch transparency, these timing paths may pass multiple latches and therefore necessitate timing constraints at these latches. For example, the timing path of type B that reaches latch $l_2$ in Fig. 2 generates setup time constraints at $l_1$ and $l_2$. The challenges in extracting these timing constraints come from the unknown arrival times at the circuit inputs during timing model extraction, and from the probabilistic latch transparency due to process variations. To solve these problems, we extract two sets of constraints, which form a lower bound and an upper bound of the real constraints from the circuit inputs, respectively. The definition and reasoning of these two constraint sets will cover most of the discussions in the rest of this section, including detailed proofs and accuracy estimation.

Consider a timing path from input $i$ to an internal latch $j$ with $n_l$ latches, where the first $n_l - 1$ latches are transparent. Similar to (12), the arrival time at the last latch $j$ on this path can be described as

$$a_j = \tilde{a}_i + \sum_{k=0}^{n_l-1}(d_{k,k+1} - E_{k,k+1}) \leq T - s_j \quad (14)$$

where we assign 0 to the index of the input $i$; and the phase shift between $i$ and the first latch on the path is assumed 0, that is, $E_{0,1} = 0$, for consistency. $\tilde{a}_i$ is the arrival time at the input $i$, which is unknown during timing model extraction. In addition, all the delays $d_{k,k+1}$ are statistical. Consequently, we can not determine how many latches a signal at the input $i$ can pass transparently during timing model extraction. In other words, we can not determine the maximum $n_l$ in (14).

*4) Double bounds of timing constraints from primary inputs*

To solve the problem of unknown transparency level, we assign a fixed maximum level of transparency $N_t$, and extract a lower bound constraint set $\underline{C}$ and an upper bound constraint set $\overline{C}$. Consider a path with more than $N_t$ latches from the input. To extract the lower bound constraint set $\underline{C}$, we only extract $N_t$ constraints in the form of (14) for the first $N_t$ latches, with $n_l = 1, \ldots N_t$ in (14). For each constraint with $n_l \geq 2$, the first $n_l - 1$ latches on the path are considered as transparent. We then discard all the constraints at latches beyond the first $N_t$ latch stages. For example, if we set $N_t$ to 2 only the setup time constraints at latch 1 and 2 in Fig. 6 are extracted. The constraints at latch 3 and further latches, even though they may be reachable from the input, are discarded.

We explain the reason that the constraint set $\underline{C}$ is a lower bound set of the real constraints using the example in Fig. 6 by setting $N_t = 2$. In the first case, consider that latch 1 is nontransparent in reality, that is, $a_1 \leq r_1$. In addition to the constraint at latch 1, we have included an unnecessary constraint in $\underline{C}$ at latch 2 by assuming $N_t = 2$. According to (14), this constraint can be written as

$$a_2 = \tilde{a}_0 + (d_{0,1} - E_{0,1}) + (d_{1,2} - E_{1,2}) \quad (15)$$
$$= a_1 + d_{1,2} - E_{1,2} \leq T - s_2. \quad (16)$$

Because the minimum clock period constraint (13) guarantees that the arrival time of the signal starting from the rising clock edge of any latch must meet the setup time constraints at any following latches [32], [40], [41], as shown in (12), we have $r_1 + d_{1,2} - E_{1,2} \leq T - s_2$. Combining this condition with $a_1 \leq r_1$ at latch 1, we can prove that (16) always holds. This shows that the redundant constraints are dominated by other constraints in the set $\underline{C}$ and do not affect the result of full-chip analysis. Therefore, with the redundant constraints described above we still have $\underline{T} = T_c$, where $\underline{T}$ is the statistical minimum clock period determined by $\underline{C}$ and $T_c$ is the one from real circuit implementation. Later in this section we will explain how to reduce the redundancy in the extracted constraint sets to improve model efficiency. In the second case, consider that latch 2 is transparent, but we do not include the timing constraints at latches beyond latch 2 in $\underline{C}$. For example, the timing constraint at latch 3 is discarded during timing model extraction. With these fewer, in other words, relaxed, timing constraints in $\underline{C}$, the computed statistical minimum clock period $\underline{T}$ during full-chip analysis is no larger than the real minimum clock period $T_c$, that is, $\underline{T} \leq T_c$. According to the analysis of the two cases above, we can conclude that $\underline{T}$ is always no larger than $T_c$, and $\underline{C}$ is a lower bound constraint set.

Now we explain the extraction of the upper bound constraint set $\overline{C}$. This set contains all the constraints in $\underline{C}$, that is, $\underline{C} \subseteq \overline{C}$. In addition, we assume that the latch at the $N_t$ stage, which is called a *boundary latch*, to be nontransparent and add the corresponding condition to $\overline{C}$. This boundary condition is written as

$$a_j = \tilde{a}_i + \sum_{k=0}^{N_t-1} (d_{k,k+1} - E_{k,k+1}) \leq r_j \quad (17)$$

where $j$ denotes the boundary latch.

In the example of Fig. 6 the boundary condition is $a_2 \leq r_2$ if we set $N_t = 2$. In the case that latch 1 or 2 in Fig. 6 is nontransparent in real circuit implementation, the necessary setup time constraints in the form of (14) have been included in $\overline{C}$, because $\underline{C} \subseteq \overline{C}$. With the redundant condition (17), the statistical minimum clock period $\overline{T}$ determined by $\overline{C}$ must be no smaller than the real minimum clock period $T_c$, that is, $T_c \leq \overline{T}$. In the other case, where the first two latches are transparent, we do not include the constraints at latches beyond latch 2 in $\overline{C}$. However, these ignored constraints are dominated by those in $\overline{C}$. For example, we ignore the constraint $a_3 = a_2 + d_{2,3} - E_{2,3} \leq T - s_3$ at latch 3 in $\overline{C}$. Similar to the discussion on $\underline{C}$, the minimum clock period constraint (13) guarantees that $r_2 + d_{2,3} - E_{2,3} \leq T - s_3$. Because we have include the nontransparent constraint $a_2 \leq r_2$ at the boundary latch 2 in $\overline{C}$. We can prove that $a_3 = a_2 + d_{2,3} - E_{2,3} \leq r_2 + d_{2,3} - E_{2,3} \leq T - s_3$ always holds if the constraints in $\overline{C}$ hold. This means that the constraints from the latches beyond the boundary latch are dominated by the constraints in $\overline{C}$, and therefore, $T_c \leq \overline{T}$. From these discussions, we can conclude that $\overline{C}$ is an upper bound constraint set.

*5) Clock scheme for extracting timing constraints*

Next, we explain the clock scheme used in the timing model extraction for latch-controlled circuits. With this clock scheme, we can estimate the accuracy of the bounding clock periods $\underline{T}$ and $\overline{T}$ determined by $\underline{C}$ and $\overline{C}$, respectively. According to the definitions of $\underline{C}$ and $\overline{C}$, we need to extract the timing constraints across latch paths with no more than $N_t$ latches. If we assume that all the parameters of the clock waveforms, for example, $r_i$, $r_j$ and $E_{ij}$ in Fig. 5, should be set by chip designers, we need to keep a lot of constraints in the timing model, leading to computational inefficiency. In practice, however, many of these parameters are predetermined and the module is optimized accordingly. Therefore, existing methods [18], [19] assume that the complete clock waveforms should be fixed when extracting the static timing model. This assumption reduces the size of the timing model and the extraction effort, but it limits the application of the timing model because only the yield at the given clock period can be evaluated. In this paper, we use a clock scheme to extract a more flexible timing model. In this clock scheme, we only assume the positions of the enabling clock edges and the phase shifts to change proportionally with the clock period assigned to the module. The period $T$ of the clock signal, however, is allowed to change in different designs. This clock scheme can be described as

$$E_{ij} = \epsilon_{ij} T \quad (18)$$
$$r_i = \epsilon_i T \quad (19)$$

where $\epsilon_{ij}$ and $\epsilon_i$ are constants with $\epsilon_{ij} > 0$ and $1 > \epsilon_i > 0$. The assumptions (18) and (19) require that the shapes of the clock signals should be scaled proportionally when the clock period $T$ is changed. This is reasonable because IP modules are typically designed and optimized with, at least some, knowledge of the clock signals in multiple-phase designs.

*6) Accuracy of the double bounds*

With the clock scheme defined in (18) and (19), we now discuss the accuracy of the two bounding clock periods $\underline{T}$ and $\overline{T}$. In the following, we estimate the distance of the two bounds, from which we then determine a maximum transparency level $N_t$ to maintain a proper accuracy. Consider a sample from Monte Carlo simulation. We know that all the constraints in $\underline{C}$ are shared by $\overline{C}$. If the additional nontransparency constraints in $\overline{C}$ are dominated by other shared constraints and therefore do not affect the upper bound $\overline{T}$, we can deduce that $\overline{T} = \underline{T}$. Otherwise, $\overline{T}$ is determined by the nontransparency condition at a boundary latch $j$. According to (17)–(19), we can write $\overline{T}$ as

$$\overline{T} = A/(\epsilon_j + B) \leq T \quad (20)$$

where $A = \tilde{a}_i + \sum_{k=0}^{N_t-1} d_{k,k+1}$ and $B = \sum_{k=0}^{N_t-1} \epsilon_{k,k+1}$. We can also write the minimum clock period $T_{st}$ determined by the setup time constraint (14) at the same boundary latch as

$$T_{st} = (A + s_j)/(1 + B) \leq T. \quad (21)$$

Because this setup time constraint is included in the lower bound constraint set $\underline{C}$, we have

$$T_{st} \leq \underline{T} \leq \overline{T}. \quad (22)$$

According to (20)–(22), the relative difference of $\underline{T}$ and $\overline{T}$ can be determined as

$$\frac{\overline{T} - \underline{T}}{\overline{T}} \leq \frac{\overline{T} - T_{st}}{\overline{T}} = 1 - \frac{(A + s_j)/(1 + B)}{A/(\epsilon_j + B)} \quad (23)$$
$$\leq 1 - \frac{A/(1 + B)}{A/(\epsilon_j + B)} = \frac{1 - \epsilon_j}{1 + B}. \quad (24)$$

The error limit (24) can be used to estimate how many levels of latch transparency should be allowed during timing model extraction. For example, in a clock scheme with two inverted clocks and duty cycles equal to 0.5, we have $1 - \epsilon_i = 0.5$ and $\epsilon_{ij} = 0.5$ for all the latches. If we set $N_t$ to 50, we can estimate that the minimum clock periods computed during full-chip analysis using the two bounds has about 2% accuracy. In practice, however, the duty cycles of the clocks are usually less than 0.5 so that $N_t$ can be smaller to reduce the number of extracted timing constraints.

*7) Reducing the number of constraints*

The constraint sets $\underline{C}$ and $\overline{C}$ contain the constraints from paths with up to $N_t$ latches. To maintain a proper accuracy many constraints may be extracted for the timing model. With the clock formulation in (18) and (19), we can reduce the number of these constraints by statistical dominance [39]. With (18) and (19), the constraints (14) and (17) can be transformed to

$$\tilde{a}_i + (s_j + \sum_{k=0}^{n_l-1} d_{k,k+1}) - (1 + \sum_{k=0}^{n_l-1} \epsilon_{k,k+1})T \leq 0 \quad (25)$$

$$\tilde{a}_i + \sum_{k=0}^{N_t-1} d_{k,k+1} - (\epsilon_j + \sum_{k=0}^{N_t-1} \epsilon_{k,k+1})T \leq 0. \quad (26)$$

Both constraints can be written as a general form

$$\tilde{a}_i + d_p + c_p T \leq 0 \quad (27)$$

where $d_p$ is a random variable and $c_p$ is a constant. Assume we have two constraints in this general form and they meet

$$\tilde{a}_i + d_{p_1} + c_{p_1}T \leq \tilde{a}_i + d_{p_2} + c_{p_2}T \iff d_{p_1} - d_{p_2} \leq (c_{p_2} - c_{p_1})T. \quad (28)$$

Then we can delete the first constraint because it is always dominated by the second. In case $c_{p_2} - c_{p_1} > 0$, (28) is equivalent to

$$(d_{p_1} - d_{p_2})/(c_{p_2} - c_{p_1}) \leq T. \quad (29)$$

Because the internal constraint (13) must be met during full-chip analysis, we can remove the dominated constraint if the probability

$$Prob\{(d_{p_1} - d_{p_2})/(c_{p_2} - c_{p_1}) \leq T_m\} \quad (30)$$

approximates 1, so that the number of constraints in the timing model can be reduced.

*8) Algorithm for extracting the constraints from primary inputs*

In real circuits, there may be many paths from an input across the internal latches so that it is very expensive to enumerate all the paths for the constraints in $\underline{C}$ and $\overline{C}$. In addition, the existence of latch loops, for example, $3 \to 4 \to 6 \to 7 \to 3$ in Fig. 6, makes a block-based traversal infeasible. In the proposed method, we use Bellman-Ford iterations [42, Ch. 25.3] to capture these constraints. The basic structure of this algorithm is shown in Algorithm 1.

In Algorithm 1, we first set the arrival time at the input $i$ to 0, because it is shared by all constraints in the form of (27) and has no effect on the constraint extraction. In each iteration, the arrival times of internal latches are updated from their fanin nodes in L4–L6. Because the arrival times in the general form on the left side of (27) may have different coefficients of T, that is, $c_p$ in (27), from different latch paths, the arrival times $a_k^{m-1}$ and $a_j^m$ in Algorithm 1 are actually sets with members indexed by $c_p$. The update in L6 computes the new arrival times from all the members in $a_k^{m-1}$ and compresses them using (28)–(30). If the node $j$ has at least one fanin node whose arrival time in the last iteration is updated, $a_j^m$ become valid in this iteration after the update in L6.

The update process described above visits the nodes in the latch graph with an event-driven mechanism. For example, in

**Algorithm 1**: Timing constraints from input $i$

| | |
|---|---|
| L1 | $\tilde{a}_i^0 = 0$; set $\tilde{a}_i^0$ as updated |
| L2 | **for** $m = 1$ **to** $N_t$ **do** |
| L3 |   **foreach** *node $j$ in the latch graph* **do** |
| L4 |     **foreach** *fanin node $k$ of $j$* **do** |
| L5 |       **if** $a_k^{m-1}$ *is updated* **then** |
| L6 |         $a_j^m \leftarrow \max_{k \to j}\{a_k^{m-1} + d_{kj} - E_{kj}\}$ |
| L7 |     **if** $a_j^m$ *is valid* **then** |
| L8 |       **if** $j$ *is a latch* **then** |
| L9 |         $extract\_st\_cons(a_j^m, s_j, T_m)$ |
| L10 |         **if** $m = N_t$ & *upper bound model* **then** |
| L11 |           $extract\_bnd\_cons(a_j^m, r_j, T_m)$ |
| L12 |       **else** |
| L13 |         $extract\_D\_cons(a_j^m)$ |
| L14 |     set $a_j^m$ as updated |

Fig. 6 the currently visited node moves across the path from 0 to 4 in the first four iterations. At the fifth iteration both node 5 and 6 are visited. As the iterations proceed, node 3 is visited again at the seventh iteration if $N_t$ is large. In further iterations the latch loop $3 \to 4 \to 6 \to 7 \to 3$ may be traversed many times. But the extracted redundant constraints at these nodes are dominated by the constraints extracted when these nodes are visited at the first time, because the timing constraint between latches (13) guarantees that the arrival time at a node on the loop does not increase after the loop is traversed [32], [41]. The basic structure of Algorithm 1 is actually a variant of Bellman-Ford algorithm and [42, Ch. 25.3] can be consulted for further details.

In Algorithm 1 the setup time constraints $a_j^m \leq T - s_j$ in the form of (14) from the arrival time set at a latch are extracted in the function $extract\_st\_cons(a_j^m, s_j, T_m)$. If an upper bound model is required and the current iteration number $m$ is equal to $N_t$, the nontransparency constraint $a_j^m \leq r_j$ is extracted in the function $extract\_bnd\_cons(a_j^m, r_j, T_m)$. In both extractions, the statistical compression technique (28)–(30) is used to reduce the number of timing constraints in $\underline{C}$ and $\overline{C}$. The condition in L8 guarantees that we only extract timing constraints at latches. These constraints are of type B in Fig. 2. If this condition is not true, the current node $j$ is an output. In this case, the update in L4–L6 computes the delays from input $i$ to output $j$ across latches and generates the constraints of type D in Fig. 2. Similar to the representation from (25) to (27), a constraint of this type defines the relation between the arrival times at input $i$ and output $j$, as

$$\tilde{a}_i + d_p + c_p T \leq a_j \quad (31)$$

where $d_p$ is a random variable and $c_p$ is a constant.

*9) Timing constraints from internal latches to primary outputs*

The last task of timing model extraction is to capture the delays from internal latches to outputs. These are the constraints of type C in Fig. 2. These delays are needed to verify the timing constraints in the following modules. Because the timing constraints are verified with no more than $N_t$ latch stages, only the delays across paths with fewer than $N_t$ latches are needed for the timing model. The extraction algorithm is a simplified version of Algorithm 1, in which the extraction of timing constraints in L8–L11 is omitted and the arrival times propagated from internal latches to the output are added to the delay sets with statistical compression. The extracted delays to output $j$ have a general form as

$$d_p + c_p T \leq a_j \tag{32}$$

which is similar to (31) except that no arrival times at inputs are involved.

## IV. HIERARCHICAL STATISTICAL TIMING ANALYSIS USING TIMING MODELS

In this section, we explain the reconstruction of spatial correlation between modules during full-chip analysis based on [43], which extends the method in [11] for a unique solution. Thereafter, we investigate how to evaluate the circuit performance using the extracted timing models.

### A. Correlation Handling with Variable Substitution

During timing model extraction, we rely on standard statistical engines to compute arrival times. In these engines gate delays are modeled using functions of independent random variables, as shown in (5). These independent random variables $\mathbf{x}$ are decomposed from the correlated variables assigned to the grid cells of the module shown in Fig. 1. Consequently, gate delays and therefore timing constraints only contain the correlation between the grid cells inside the module. For example, the timing models of the two modules in Fig. 7 contain the statistical constraints represented by their individual independent components.

To reconstruct the correlation between different modules, we first transform the variables in an extracted timing model back to the variables assigned for the grid cells of the module. Assume we have a timing constraint from module A. The corresponding random variables, for example, an edge delay in the compressed timing graph, can be written as (5). According to the decomposition (2), we can transform $\mathbf{x}$ back to $\mathbf{p}_l$ by

$$\mathbf{x} = \mathbf{A}^T \mathbf{p}_l \tag{33}$$

where $\mathbf{A}$ is the orthogonal transformation matrix for decomposing the $n$ correlated parameters $\mathbf{p}_l$ of module A. Thereafter, the delay $d$ becomes

$$d = g(\mathbf{k}\mathbf{A}^T \mathbf{p}_l). \tag{34}$$

With this transformation, we represent all the variables in the timing model as functions of the correlated random variables originally assigned to the grid cells of the module.

The key idea of the correlation reconstruction is to partition each module with the same grid as during timing model extraction. For example, the partition of module A in Fig. 7 should be the same as that in Fig. 1. All the grid cells of the modules in Fig. 7 together form the grid partition at chip level. For example, the die partition in Fig. 7 has 15 grid cells. Similar to modeling the correlation described in Section II-A, we assign a variable to each of these grid cells. Assume there are in total $m$ grid cells at chip level and the $m$ random variables assigned to the grid cells are written as a vector $\mathbf{p}_l^t$. By applying variable decomposition similar to (2), $\mathbf{p}_l^t$ can be written as

$$\mathbf{p}_l^t = \mathbf{B}\mathbf{x}^t \tag{35}$$

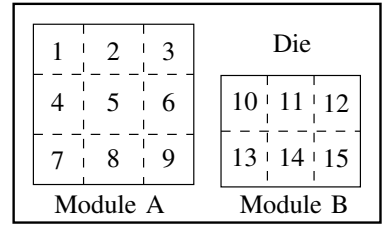

Fig. 7. Grid partition of the chip.

where $\mathbf{B}$ is the orthogonal transformation matrix. $\mathbf{x}^t$ is the vector of independent random variables.

We assume there are $n$ grid cells in the area of module A and write the corresponding variables as a vector $\mathbf{p}_{l,n}^t$. These random variables are part of $\mathbf{p}_l^t$ which contains the random variables assigned to all the grid cells in the chip. Because (34) simply means that $d$ is a function of the random variables assigned to the grid cells of module A, we can write $d$ as

$$d = g(\mathbf{k}\mathbf{A}^T \mathbf{p}_{l,n}^t). \tag{36}$$

According to (35), $\mathbf{p}_{l,n}^t$ can be decomposed as

$$\mathbf{p}_{l,n}^t = \mathbf{B}_n \mathbf{x}^t \tag{37}$$

where the $n \times m$ matrix $\mathbf{B}_n$ contains the $n$ rows of the transformation matrix $\mathbf{B}$ corresponding to $\mathbf{p}_{l,n}^t$, which is a subvector of $\mathbf{p}_l^t$ in (35). From (36) and (37) we can finally transform $d$ as

$$d = g(\mathbf{k}\mathbf{A}^T \mathbf{B}_n \mathbf{x}^t). \tag{38}$$

According to (38) and (5), we can transform all the independent components in the timing models to the independent components at chip level using the corresponding transformation matrices by

$$\mathbf{x} = \mathbf{A}^T \mathbf{B}_n \mathbf{x}^t \tag{39}$$

where $\mathbf{B}_n$ is selected from $\mathbf{B}$ and $\mathbf{A}$ is the transformation matrix for each individual module. After this transformation, all timing constraints in the timing models become functions of the same independent components $\mathbf{x}^t$ and the spatial correlation is represented by sharing the same set of random variables.

In variable decomposition, unimportant components may be discarded. For example, the eigenvalues as well as the corresponding eigenvectors of the correlation matrix after PCA are sorted in descending order. The variables with small eigenvalues can be discarded without affecting the accuracy of the decomposition substantially. Assume we select the first $n'$ components from $\mathbf{x}$ and $m'$ components from $\mathbf{x}^t$. The last $n - n'$ rows in $\mathbf{A}^T$ and $m - m'$ columns in $\mathbf{B}_n^t$ can be considered as zero during the transformation. Because usually the number of selected components is much smaller than the number of the original variables, this simplification can significantly reduce the effort in computing the multiplication of $\mathbf{A}^T \mathbf{B}_n$. With this transformation, we can avoid the variable selection problem in [11] and guarantee a unique solution in the variable substitution.

### B. Timing Analysis using Extracted Models

Timing models for combinational circuits are only compressed timing graphs, so that standard statistical timing engines can be used directly for full-chip analysis. When timing models of flip-flop-based circuits are used, the arrival times at the inputs of a module can be computed as the maximum delays from the fanin modules. Thereafter, the timing constraints of the inputs, together with the minimum clock periods of the modules in the form of (10), can be used to evaluate the statistical performance of the complete design.

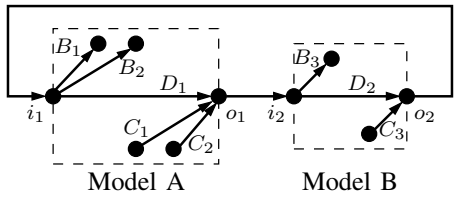

Fig. 8. Hierarchical timing constraints with latch transparency.

In timing models of latch-controlled circuits, the constraints of type D in Fig. 2 establish connections between inputs and outputs. These constraints may form loops at chip level and therefore make the timing analysis more complicated, because timing verification may traverse the loops many times. We show an example of the chip-level structure of the timing constraints from latch-controlled modules in Fig. 8. This structure is called *constraint graph* in the following. In this example, the circuit contains two modules, each of which has one input and one output. Timing model A contains two constraint edges, $B_1$ and $B_2$, at its input, representing the timing constraints of type B in Fig. 2, in the form of (27); and two delay edges, $C_1$ and $C_2$, from its internal latches to its output, representing the timing constraints of type C in Fig. 2, in the form of (32). Timing model B contains only one constraint edge $B_3$ and one delay edge $C_3$. The edges $D_1$ and $D_2$ in these timing models represent the timing constraints of type D in Fig. 2, in the form of (31).

In Fig. 8 the timing constraints and chip-level connections form a loop, which makes a block-based analysis infeasible. In this paper, we use a fast method compatible with the upper and lower bounds described in Section III-C to evaluate the circuit performance. The basic concept of this algorithm is shown in Algorithm 2. For convenience of explaining this algorithm, we call the nodes at the end of edges $B_1$, $B_2$ and $B_3$ in Fig. 8 *constraint nodes*; the nodes at the beginning of edges $C_1$, $C_2$ and $C_3$ *delay nodes*; and the other nodes *input/output nodes*.

To guarantee the proper behavior of the circuit, the arrival times of the signals starting from delay nodes must be verified at all reachable constraint nodes. To avoid traversing across loops infinitely, we force nontransparency at the constraint nodes where no fewer than $N_t$ latch stages are traversed. This technique is shown in L5–L10 in Algorithm 2, where $update\_arr(a_j^m)$ corresponds to L4–L6 in Algorithm 1. In Algorithm 2, the timing constraints from modules are verified using $verify\_cons(a_j^m)$. These include the constraints guaranteeing setup time (14) from $\underline{C}$ and $\overline{C}$ and the constraints forcing the nontransparency condition (17) from $\overline{C}$. According to the general form (27), each of these constraints defines a lower bound for the clock period of the circuit, that is,

$$(\tilde{a}_i + d_p)/(-c_p) \leq T \qquad (40)$$

where $c_p$ is always negative. $\tilde{a}_i$ is the arrival time at input $i$ of the module whose value is updated during the iterations. The minimum clock period of the circuit is computed by the maximum of these lower bounds and $T_m$ in (13). After this extraction, we then check whether the level of traversed latch stages exceeds $N_t$. In case this is true, we force a nontransparency condition so that the propagation of arrival times across this module can be broken. This extraction is performed by the function $verify\_nontrans(a_j^m)$. With the newly created nontransparency constraints, the arrival times at input nodes can be deleted if the level of latch stages exceeds $N_t$, because under this condition the signals with these arrival times must reach a nontransparent latch inside the module and can not generate further constraints. This arrival time removal is shown in L11–L13. The last part of Algorithm 2, L14–L17, updates the arrival times at the input/output nodes for further iterations. Because the edges from inputs to outputs in Fig. 8 are formed across multiple latch stages, the level of latch transparency across modules increases very fast. Therefore using L11–L13 we can reduce the arrival times effectively during the iterations and the algorithm converges very fast.

**Algorithm 2**: Hierarchical verification with loops

| | |
|---|---|
| L1 | **foreach** *delay node $j$ in the constraint graph* **do** |
| L2 |    $a_j^0 = 0$; set $a_j^0$ as updated |
| L3 | $m = 1$ |
| L4 | **while** *at least a node is updated* **do** |
| L5 |    **foreach** *constraint node $j$* **do** |
| L6 |       $update\_arr(a_j^m)$ |
| L7 |       $verify\_cons(a_j^m)$ |
| L8 |       **if** $latch\_stage(a_j^m) \geq N_t$ **then** |
| L9 |          **if** *verifying upper bound* **then** |
| L10 |             $verify\_nontrans(a_j^m)$ |
| L11 |    **foreach** *input node $j$* **do** |
| L12 |       **if** $latch\_stage(a_j^{m-1}) \geq N_t$ **then** |
| L13 |          clear $a_j^{m-1}$ |
| L14 |    **foreach** *input/output node $j$* **do** |
| L15 |       $update\_arr(a_j^m)$ |
| L16 |       **if** $a_j^m$ *is valid* **then** |
| L17 |          set $a_j^m$ as updated |
| L18 |    $m = m + 1$ |

## V. Experimental Results

The proposed method was implemented in C++. The computer used for experiments has a CPU running at 2.66 GHz with four cores and 12 Gb memory. The operating system is Linux. The proposed method has a single-threaded implementation. The Monte Carlo simulations were distributed to multiple cores, but we subtracted the runtime required for task distribution from the overall time consumption, so that we can compare only the net runtime of Monte Carlo simulations with the runtime of our proposed method.

The cells in the benchmark circuits were mapped to a library from an industry partner. The standard deviations of transistor length, oxide thickness and threshold voltage were assigned to 15.7%, 5.3% and 4.4% of the nominal values, respectively [44]. The cell delays were created using the method proposed in [1]. Five correlation curves shown in Fig. 9 were used to model the spatial correlation of process variations. These curves were adapted from [26] and have similar shapes as the correlation measurements shown in [24]. In these curves E has the strongest correlation with respect to on-die distance in grid cells. We used curve C in most of our experiments. The other curves are used only to test the necessity of applying the variable substitution in Section IV-A. In the experiments, we used the SSTA engines proposed in [2] and [45] to compute the sum and maximum of random variables and the method in [34] to compute the criticality $c_{ij}^{pq}$ in (6). The results from hierarchical statistical timing analysis were compared with those from Monte Carlo simulations with 10 000 samples to verify the accuracy of the proposed method.

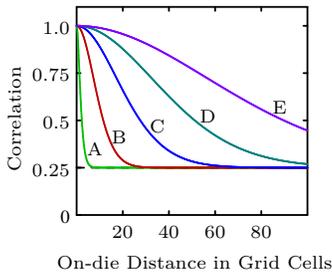

Fig. 9. Correlation curves.

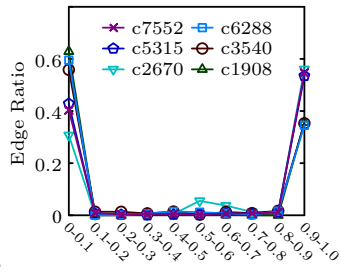

Fig. 10. Criticality distribution.

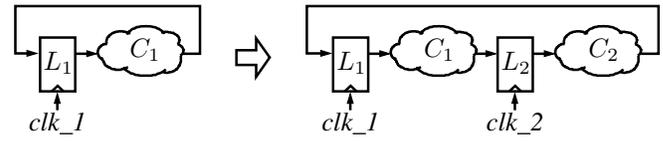

Fig. 11. Circuit construction for sequential modules.

For testing combinational timing models, we used ISCAS85 benchmark circuits as modules. For sequential models with flip-flops and latches, we constructed larger modules from the ISCAS89 benchmark circuits by duplicating the registers $L_1$ and combinational circuits $C_1$ as illustrated in Fig. 11, which is adapted from [46, Fig. 6]. These constructed circuits are used in [46] for testing algorithms of static timing analysis of latch-controlled circuits. We refer to these constructed circuits by adding the prefix '2' to the names of the original ISCAS89 benchmark circuits. When testing latch-controlled modules, the clocks $clk\_1$ and $clk\_2$ formed a two-phase clock scheme and were set to inverted clocks with clock phase shift $T/2$ and duty cycle equal to 0.5. For convenience, we also used these circuits to test timing models for flip-flip-based modules, where the registers were simply assumed as edge-triggered. We built hierarchical designs with $n$ rows and $n$ columns of the modules described above. For combinational modules we set $n = 6$ and for sequential modules $n = 2$, so that the largest designs in these experiments are comparable in size. The connections between these modules were generated randomly.

### A. Results of Combinational Models

The timing model extraction of combinational modules in Section III-A depends on the percentage of noncritical edges in the timing graph. To test this property, we assign the edges in each of the six benchmark circuits into ten groups according to their criticalities defined in (6), and the ratios of the edges in these groups are illustrated in Fig. 10, where the y-axis shows the ratios and x-axis shows the criticality groups. In extracting timing models gate delays are critical if they are on any critical path with respect to any input-output pair. With this definition, gate delays have a higher chance to be critical in timing model extraction than in circuit optimization, so that their criticalities have the tendency to be 1, and only the edges which will never be critical to the timing model have criticalities approximating 0. Therefore the overall criticality distributions have the tendency shown in Fig. 10 when all delays are considered. In the timing model extraction, we set the criticality threshold $\delta_c$ in Section III-A to 0.1, so that the extracted timing models preserve the critical edges to maintain a proper accuracy. The results of these timing models are shown in Table I. In this table, $n_i$ and $n_o$ are the numbers of inputs and outputs of a module, respectively. $n_c$ is the number of combinational gates. $\mathcal{R}_e$ is the ratio of the number of edges in the timing model to the number of edges in the original module. $\mathcal{R}_v$ is the ratio of the numbers of nodes. From these ratios, we can find that the extracted timing models are much smaller than the original modules. Especially the benchmark circuit c6288 has very small ratios, because this circuit contains very long paths which have a higher chance to dominate other paths.

Because a combinational timing model should preserve the delays from inputs to outputs of the module, we have computed these delays from the extracted timing models and compared them with those from Monte Carlo simulations using the original modules to verify the accuracy. In Table I $\mathcal{E}_\mu^m$ is the maximum relative error of the means of all input-output delays and $\mathcal{E}_\sigma^m$ is the maximum relative error of standard deviations of these delays. From these comparisons, we can confirm that the accuracy of the extracted timing models is still maintained. The runtimes of extracting timing models are shown as $T_e$ in Table I. With these runtimes we can confirm the efficiency of the proposed method.

In Table I we also show the results of hierarchical timing analysis of the circuits constructed by 6 rows and 6 columns of the modules. We first computed the maximum delay of the circuit using the extracted models. Then we compared the mean of this delay and that from Monte Carlo simulation using the flattened circuit. The relative error of this mean value is shown as $\mathcal{E}_\mu^t$ in Table I. Similar to $\mathcal{E}_\mu^t$, $\mathcal{E}_\sigma^t$ is defined to show the accuracy of the standard deviation. For comparison, we also computed the maximum delay of the circuit using the statistical timing engine with all modules flattened. The relative errors are shown as $\mathcal{E}_\mu^f$ and $\mathcal{E}_\sigma^f$ in Table I. By comparing these relative errors, we can find that statistical timing analysis with the extracted timing models maintains a similar accuracy as the flattened analysis.

The runtimes of the variable substitutions are shown as $T_v$ in Table I. The runtimes of statistical timing analysis using timing models after variable substitution are shown as $T_r$ and the runtimes of flattened analysis as $T_f$. For comparison, the runtimes of Monte Carlo simulations using the flattened circuits are also listed as $T_{MC}$ in Table I. Comparing $T_v + T_r$ with $T_f$ we can see that using timing models the runtimes are much shorter than using the flattened circuits. In practice, the variable substitution is very often performed only once due to floorplan, especially when incremental analysis is considered. In such cases, the acceleration of using the extracted timing

TABLE I
RESULTS OF COMBINATIONAL MODELS

| Circuit | $n_i$ | $n_o$ | $n_c$ | Combinational Timing Models | | | | | Analysis with Timing Models | | | | Flattened Analysis | | | |
|---|---|---|---|---|---|---|---|---|---|---|---|---|---|---|---|---|
| | | | | $\mathcal{R}_e(\%)$ | $\mathcal{R}_v(\%)$ | $\mathcal{E}_\mu^m(\%)$ | $\mathcal{E}_\sigma^m(\%)$ | $T_e(s)$ | $\mathcal{E}_\mu^t(\%)$ | $\mathcal{E}_\sigma^t(\%)$ | $T_v(s)$ | $T_r(s)$ | $\mathcal{E}_\mu^f(\%)$ | $\mathcal{E}_\sigma^f(\%)$ | $T_f(s)$ | $T_{MC}(s)$ |
| c432 | 36 | 7 | 160 | 13.39 | 23.47 | 0.02 | 0.75 | 0.02 | 0.67 | 0.98 | 0.01 | 0.01 | 1.08 | 0.59 | 0.03 | 154.46 |
| c499 | 41 | 32 | 202 | 42.16 | 40.74 | 0.06 | 1.08 | 0.02 | 0.74 | 1.39 | 0.02 | 0.02 | 0.68 | 1.08 | 0.05 | 220.51 |
| c880 | 60 | 26 | 383 | 33.06 | 25.73 | 0.05 | 0.41 | 0.02 | 0.70 | 1.20 | 0.03 | 0.02 | 0.62 | 0.69 | 0.09 | 362.61 |
| c1355 | 41 | 32 | 546 | 13.35 | 16.87 | 0.13 | 0.90 | 0.08 | 0.64 | 0.78 | 0.02 | 0.01 | 0.38 | 0.69 | 0.13 | 669.33 |
| c1908 | 33 | 25 | 880 | 12.82 | 9.31 | 0.03 | 0.97 | 0.08 | 1.40 | 0.08 | 0.04 | 0.02 | 1.65 | 0.86 | 0.20 | 835.35 |
| c2670 | 233 | 140 | 1193 | 19.46 | 23.49 | 0.17 | 0.34 | 0.07 | 0.70 | 0.99 | 0.12 | 0.07 | 0.59 | 0.39 | 0.30 | 1368.45 |
| c3540 | 50 | 22 | 1669 | 13.03 | 7.27 | 0.08 | 0.84 | 0.13 | 1.16 | 0.11 | 0.11 | 0.03 | 1.08 | 1.02 | 0.43 | 1770.37 |
| c5315 | 178 | 123 | 2307 | 21.43 | 16.70 | 0.22 | 0.60 | 0.16 | 0.12 | 1.01 | 0.44 | 0.12 | 0.07 | 1.57 | 0.80 | 3013.97 |
| c6288 | 32 | 32 | 2416 | 9.94 | 8.66 | 0.22 | 0.52 | 0.49 | 1.17 | 1.58 | 0.34 | 0.05 | 0.54 | 1.62 | 0.82 | 3077.01 |
| c7552 | 207 | 108 | 3512 | 17.12 | 14.65 | 0.57 | 0.77 | 0.25 | 0.28 | 1.05 | 0.64 | 0.14 | 0.43 | 1.13 | 1.22 | 4023.80 |

TABLE II
RESULTS OF FLIP-FLOP-BASED MODELS

| Circuit | $n_i$ | $n_o$ | $n_c$ | $n_s$ | $T_e(s)$ | Analysis with Timing Models | | | | Flattened Analysis | | | |
|---|---|---|---|---|---|---|---|---|---|---|---|---|---|
| | | | | | | $\mathcal{E}_\mu^t(\%)$ | $\mathcal{E}_\sigma^t(\%)$ | $T_v(s)$ | $T_r(s)$ | $\mathcal{E}_\mu^f(\%)$ | $\mathcal{E}_\sigma^f(\%)$ | $T_f(s)$ | $T_{MC}(s)$ |
| 2s298 | 3 | 6 | 238 | 28 | <0.01 | 0.78 | 0.51 | <0.01 | <0.01 | 0.46 | 0.47 | <0.01 | 12.33 |
| 2s526 | 3 | 6 | 386 | 42 | <0.01 | 0.25 | 0.79 | <0.01 | <0.01 | 0.86 | 1.30 | 0.01 | 34.01 |
| 2s820 | 18 | 7 | 2312 | 40 | 0.01 | 0.00 | 0.95 | <0.01 | <0.01 | 1.11 | 0.53 | 0.01 | 37.82 |
| 2s1238 | 14 | 2 | 1016 | 36 | 0.02 | 0.05 | 0.77 | <0.01 | <0.01 | 0.18 | 0.13 | 0.01 | 20.71 |
| 2s1423 | 17 | 2 | 1314 | 148 | 0.03 | 0.37 | 1.38 | <0.01 | <0.01 | 0.57 | 1.33 | 0.04 | 138.91 |
| 2s5378 | 35 | 35 | 5558 | 358 | 0.12 | 0.35 | 1.22 | <0.01 | <0.01 | 0.37 | 1.45 | 0.17 | 588.21 |
| 2s9234 | 36 | 10 | 11194 | 422 | 0.23 | 0.33 | 0.67 | 0.01 | <0.01 | 0.06 | 0.42 | 0.28 | 1175.79 |
| 2s13207 | 62 | 100 | 15902 | 1276 | 0.37 | 0.75 | 0.82 | 0.05 | <0.01 | 0.34 | 0.04 | 0.48 | 2173.84 |
| 2s15850 | 77 | 41 | 19544 | 1068 | 0.67 | 0.04 | 0.10 | 0.07 | <0.01 | 0.33 | 0.93 | 0.62 | 2406.68 |
| 2s38584 | 38 | 219 | 38506 | 2852 | 1.46 | 0.17 | 1.66 | 0.58 | <0.01 | 0.35 | 0.92 | 2.27 | 7064.29 |

models is even larger when $T_r$ and $T_f$ are compared.

### B. Results of Flip-flop-based Models and Variable Substitution

The experimental results of flip-flop-based timing models are shown in Table II. Each module has $n_i$ inputs, $n_o$ outputs, $n_c$ combinational gates and $n_s$ flip-flops. According to Section III-B, a timing model for such a module contains $n_i$ constraints and $n_o$ delays. In addition, the timing model contains the minimum clock period for the internal flip-flops, shown in (10). The runtimes of extracting these timing models are listed in Table II as $T_e$, where runtimes less than 0.01 second are of no significance and not measured accurately. We tested the extracted timing models using hierarchical circuits built from 2 rows and 2 columns of the modules in Table II. In this table, $\mathcal{E}_\mu^t$ and $\mathcal{E}_\sigma^t$ are the relative errors of the minimum clock periods using the extracted timing models compared with those from Monte Carlo simulations. $\mathcal{E}_\mu^f$ and $\mathcal{E}_\sigma^f$ are relative errors of flattened analysis. With these comparisons we can confirm the accuracy of timing analysis using the extracted timing models. Because the constraints between all internal flip-flops inside the modules are compressed, the timing models are much smaller than the original circuits. By comparing the runtimes $T_v$, $T_r$ and $T_f$ in Table II, we can find that timing analysis using the extracted models is much faster than using the flattened circuits directly.

In Section IV-A we have explained the variable substitution method to reconstruct the correlation between modules. We justify the necessity of this correlation reconstruction by illustrating the accuracy of the method considering only the correlation between dies in Fig. 12. For each test case we apply different correlation curves A–E from Fig. 9, where curve A has the smallest correlation and curve E the largest. The y-axes in Fig. 12 show the errors of mean and standard deviation when spatial correlation is not considered. From these two figures, we can observe that ignoring spatial correlation between modules may cause significant loss in accuracy. Furthermore, all the curves show a similar trend, starting with a small error, reaching a peak and then descending. This common trend can be explained as follows. If the correlation decreases very fast when the distance increases, the spatial correlation has relatively weak effect on accuracy. With more correlation from curves A to E, the effect of correlation increases. However, when the correlation is very large and increases further, the devices on the die can share more correlation together, leading to a small error again when discarding spatial correlation. This explains the peaks in the accuracy curves. From Fig. 12 we can also observe that the smaller the size of the circuit is, the earlier the peak appears. This is simply because smaller circuits can share more spatial correlation than larger ones.

### C. Results of Latch-controlled Models

To test the timing models of latch-controlled circuits, the registers in the 2 rows and 2 columns of modules are assumed as level-sensitive latches with a two-phase clock scheme. The results are shown in Table III. In the experiments, we set the maximum level of transparency $N_t$ described in Section III-C to 50 to maintain a proper accuracy. Owing to latch transparency, we have more than one constraint at each input of the module. The average number of constraints at an input is shown as $n_{cons}$ in Table III. The average number of delays to an output of the module is shown as $n_{delay}$. Because many inputs may have paths across transparent latches to reach outputs, the average numbers of delays are much larger than the numbers of constraints. The runtimes to extract these

TABLE III
RESULTS OF LATCH-CONTROLLED MODELS

| Circuit | Timing Models with Latches | | | Analysis with Timing Models | | | | | | Flattened Analysis | | | |
|---|---|---|---|---|---|---|---|---|---|---|---|---|---|
| | $n_{cons}$ | $n_{delay}$ | $T_e(s)$ | $\mathcal{E}_{\underline{\mu}}(\%)$ | $\mathcal{E}_{\underline{\sigma}}(\%)$ | $\mathcal{E}_{\overline{\mu}}(\%)$ | $\mathcal{E}_{\overline{\sigma}}(\%)$ | $T_v(s)$ | $T_r(s)$ | $\mathcal{E}_\mu^f(\%)$ | $\mathcal{E}_\sigma^f(\%)$ | $T_f(s)$ | $T_{MC}(s)$ |
| 2s298 | 2.67 | 5.17 | 0.02 | -0.93 | -0.98 | 0.99 | 0.95 | <0.01 | 0.01 | 0.16 | 0.67 | 0.1 | 631.88 |
| 2s526 | 2.33 | 8.00 | 0.04 | -0.49 | -1.44 | 0.94 | 0.03 | <0.01 | 0.21 | 0.93 | 0.69 | 0.35 | 2180.64 |
| 2s820 | 1.61 | 19.86 | 0.02 | -0.74 | -0.57 | 1.19 | 1.30 | <0.01 | 0.02 | 0.21 | 1.36 | 0.03 | 332.31 |
| 2s1238 | 1.00 | 7.00 | 0.01 | -0.11 | 0.52 | -0.05 | 0.11 | <0.01 | <0.01 | 0.16 | 0.41 | 0.11 | 498.66 |
| 2s1423 | 3.24 | 19.00 | 2.06 | 0.04 | 0.72 | 0.05 | 0.68 | <0.01 | 0.01 | 0.16 | 0.26 | 17.21 | 385171.46 |
| 2s5378 | 3.33 | 44.16 | 1.65 | -0.25 | -1.31 | 0.93 | 0.69 | 0.01 | 0.61 | 1.42 | 0.69 | 8.22 | 1185787.59 |
| 2s9234† | 19.64 | 1.60 | 2.58 | -1.00 | -1.12 | 0.33 | 0.82 | 0.02 | 0.07 | - | - | 9.52 | - |
| 2s13207† | 18.85 | 79.16 | 3.45 | -1.28 | -1.13 | 0.67 | 0.71 | 0.12 | 0.74 | - | - | 123.27 | - |
| 2s15850† | 2.97 | 21.00 | 25.06 | -0.35 | -1.00 | 0.84 | -0.83 | 0.08 | 0.07 | - | - | 174.42 | - |
| 2s38584† | 16.31 | 15.51 | 16.34 | -0.60 | 0.29 | -0.34 | 0.25 | 0.59 | 0.1 | - | - | 61.91 | - |

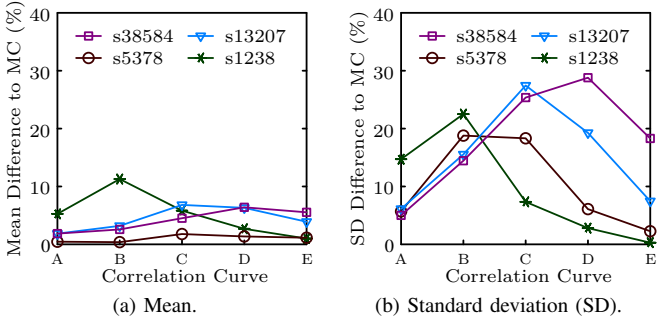

Fig. 12. Accuracy without considering spatial correlation.

(a) Mean. (b) Standard deviation (SD).

timing models are shown as $T_e$ in Table III. Even for large modules the proposed method consumes no more than 30 seconds, which is acceptable for timing model extraction.

To verify the accuracy of applying these timing models in hierarchical analysis, we have run Monte Carlo simulations with the circuits constructed from 2s298 to 2s5378. The Monte Carlo simulations for the circuits marked with † would consume months to finish and therefore are unaffordable even with parallelism in our experiments. We have tested the proposed lower- and upper-bound constraint sets described in Section III-C. The relative errors of using the lower-bound constraint set are shown as $\mathcal{E}_{\underline{\mu}}^t$ and $\mathcal{E}_{\underline{\sigma}}^t$; the relative errors using the upper-bound constraint sets are $\mathcal{E}_{\overline{\mu}}^t$ and $\mathcal{E}_{\overline{\sigma}}^t$. For the circuits marked by † the errors are computed by comparing the means and standard deviations with those from statistical timing analysis using the flattened circuits, because no corresponding results from Monte Carlo simulation are available. The relative errors of statistical timing analysis using flattened circuits compared to Monte Carlo simulations are shown as $\mathcal{E}_\mu^f$ and $\mathcal{E}_\sigma^f$. From these comparisons we can see that both the lower- and upper-bound constraint sets have accuracy comparable to statistical timing analysis using flattened circuits.

In Table III the runtimes of variable substitution are listed as $T_v$; the runtimes of hierarchical timing analysis using extracted timing models as $T_r$; and the runtimes of statistical timing analysis with flattened circuits as $T_f$. From these runtimes, we can observe that hierarchical statistical timing analysis has a dominating advantage, due to the fact that statistical timing analysis for latch-controlled circuits does not scale linearly when the size of the circuit increases.

## VI. CONCLUSION

In this paper, we proposed effective timing model extraction methods for combinational, flip-flop-based and latch-controlled circuits. The extracted timing models do not expose the internal circuit structure of IP blocks and therefore are suitable for IP protection. For full-chip timing analysis, we proposed a method based on variable substitution to reconstruct the correlation between modules. We also proposed an efficient method compatible with the lower- and upper-bound constraint sets for latch-controlled circuits in applying the extracted timing models. Experimental results confirm that the proposed hierarchical statistical design flow reduces the runtime of full-chip timing analysis significantly and the accuracy of statistical timing analysis is maintained.

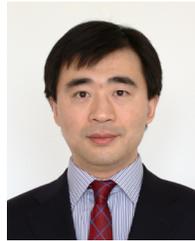

**Bing Li** received the Bachelors and Masters degrees in communication and information engineering from the Beijing University of Posts and Telecommunications, Beijing, China, in 2000 and 2003, respectively, and the Ph.D. degree in electrical engineering from Technische Universität München (TUM), Munich, Germany, in 2010.

He is currently a Post-Doctoral Researcher with the Institute for Electronic Design Automation, TUM. His current research interests include timing and power analysis of integrated circuits.

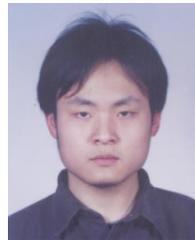

**Ning Chen** received the Dipl.-Ing. degree in electrical engineering and information technology from Technische Universität München (TUM), Munich, Germany, in 2007.

He has been a Research and Teaching Assistant with the Institute for Electronic Design Automation, TUM, since 2008. His current research interests include electronic design automation for digital circuits with a focus on timing analysis.

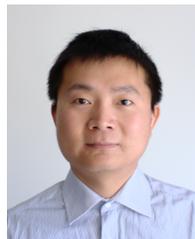

**Yang Xu** received the Bachelors degree in electronic and information engineering from Zhejiang University, Hangzhou, China, in 2005 and the Masters degree in communications engineering from Technische Universität München (TUM), Munich, Germany, in 2007.

He is currently a PhD student with Intel Mobile Communications and the Chair of Hardware-Software-Codesign in the University of Erlangen-Nuremberg, Germany. His current research interests include modeling and analysis of power and performance of integrated circuits.

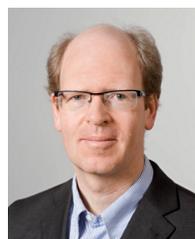

**Ulf Schlichtmann** (S'88–M'90) received the Dipl.-Ing. and Dr.-Ing. degrees in electrical engineering and information technology from Technische Universität München (TUM), Munich, Germany, in 1990 and 1995, respectively.

He was with Siemens AG and Infineon Technologies AG, Munich, from 1994 to 2003, where he held various technical and management positions in design automation, design libraries, IP reuse, and product development. He has been with TUM as a Professor and the Head of the Institute for Electronic Design Automation, since 2003. He served as the Dean of the Department of Electrical Engineering and Information Technology, TUM, from 2008 to 2011. His current research interests include computer-aided design of electronic circuits and systems, with an emphasis on designing robust systems.